# Unique opportunity to harness polarization in GaN to override the conventional power electronics figure-of-merits

Huili Grace Xing[1,2,3] , Bo Song[1,3], Mingda Zhu[3], Zongyang Hu[1,3], Meng Qi[3], Kazuki Nomoto[3] & Debdeep Jena[1,2,3]
[1]School of Electrical and Computer Engineering, Cornell University, Ithaca, NY 14853, USA
[2]Department of Materials Science and Engineering, Cornell University, Ithaca, NY 14853, USA
[3]Department of Electrical Engineering, University of Notre Dame, Notre Dame, IN 46556, USA
Email: grace.xing@cornell.edu

Owing to the large breakdown electric field ($E_b$), wide bandgap semiconductors (WBGs) such as SiC, GaN, $Ga_2O_3$ and diamond based power devices are the focus for next generation power switching applications. The unipolar trade-off relationship between the area specific-on resistance ($R_{on,sp}$) and breakdown voltage ($BV$) is often employed to compare the performance limitation among various materials. The advanced features in benchmarking power devices include conduction modulation and superjunction in devices taking advantage of bipolar conduction. To this end, the GaN material system has a unique advantage due to its prominent spontaneous and piezoelectric polarization effects in GaN, AlN, InN, $Al_xIn_yGaN$ alloys and flexibility in inserting appropriate heterojunctions thus dramatically broaden the device design space. For instance, $Al_xGaN$, which has a larger bandgap, can be designed into the highest field region. Another enabling feature in the GaN material system is polarization doping or Pi-doping: by compositionally grading $Al_XIn_yGaN$, both electron and hole doping can be achieved, exhibiting the ideal dopant behavior – no temperature dependence or frequency dependence! This dopant characteristics can't be taken for granted for WBGs based on the knowledge in Si. *Pi-doping is an extremely powerful attribute since all wide bandgap semiconductors face the challenge of deep dopants while this ideal solution exists in GaN.*

Fig.1 reviews the fundamentals on Pi-doping as well as our past work. On the Ga-face GaN, when grading from GaN to AlGaN, the difference in spontaneous and piezoelectric polarization charges in $Al_xGaN$ to $Al_{x+\delta}GaN$ leads to a 3-dimensionally distributed positive charge in the crystal layer (Fig. 1) [1]. These positive charges associated with the crystal lattice are immobile, in turn attracting mobile negative charge from the entire material (often surface donors) to minimize the electric field within the crystal. To obtain holes, one just needs to grading back from $Al_xGaN$ to GaN on Ga face or grading from GaN to $Al_xGaN$ on N face [2]. Its major difference from the conventional impurity induced doping is that the polarization-induced electrons (or holes) are *electric field ionized* but the impurity-doped electrons (holes) result from thermal excitation of electrons (holes) at the donor (acceptor) energy states to the conduction (valence) energy band of the semiconductor. As a result, the polarization-induced electrons do not suffer from carrier freeze-out as the impurity-doped electrons do when the temperature decreases. Meanwhile, the polarization-doped electrons typically exhibit higher mobilities due to the absence of ionized impurity scattering [3]. Fig.2 shows the charge density as a function of Al composition $x$ and grading layer thickness $d$ when linearly grading GaN to $Al_xGaN$, along with the reported data to date. The past research has been mainly focused on high carrier concentrations $>10^{18}cm^{-3}$, while for power switching applications doping concentrations below $10^{17}$ $cm^{-3}$ are desired for high breakdown voltage. Thus, a wide-open and unique research opportunity has emerged to explore polarization engineering for power electronics.

To better illustrate unique technologies in GaN, two examples are presented here. *The first device is PolarMOS (Fig.3) [4]*. The device structure consists of a drift n-GaN region (by impurity and Pi-doping), a Pi-doped p-type $Al_xGaN$, a thin GaN channel followed by an AlGaN barrier. It features a high threshold $V_{th}$> 2 V by employing a gate dielectric layer. The vertical n-GaN channel can be formed by either converting a Pi-doped p-layer to n-type by ion implantation or selective-area epitaxial regrowth. An added benefit of a p-$Al_{0\to x}GaN$/n-$Al_{x\to 0}GaN$ junction is that the p-n junction depletion falls in the AlGaN layer, where a higher critical electrical field can be supported. Our simulation results suggest that the PolarMOS promise a higher $V_{br}^2/R_{on,sp}$ thus standing out compared to other solutions in the same neighborhood of the voltage-current ratings. *The second device is a lateral polarization-doped super junction (LPSJ) (Fig. 4) [5-6]*, which consists of epitaxially grown n/p pillar regions realized by compositionally grading AlGaN, followed by a regrowth of the $n^+$ and $p^+$ cathode/anode regions connecting with the n and p pillars [7,8]. To realize the super junction, i.e. n/p pillars with balanced charges, one only needs to linearly grade from GaN to $Al_xGaN$ and then grading back to GaN during epitaxial growth, without an ion implantation process used in Si. The doping profile of the pillars can be precisely controlled by MBE/MOCVD growth in terms of Al composition and layer thicknesses, thus suppressing the charge imbalance problem. Extensive modeling shows that LPSJ can not only outperformance the state-of-the-art devices, but also break the unipolar limit of GaN material, where $R_{on,sp}$ of LPSJ with $x_{Al}$ = 0.3 shows > 10X reduction over conventional GaN junctions for BV>2 kV.

Acknowledgement: this work is in part supported by the ARPAe SWITCHES project.

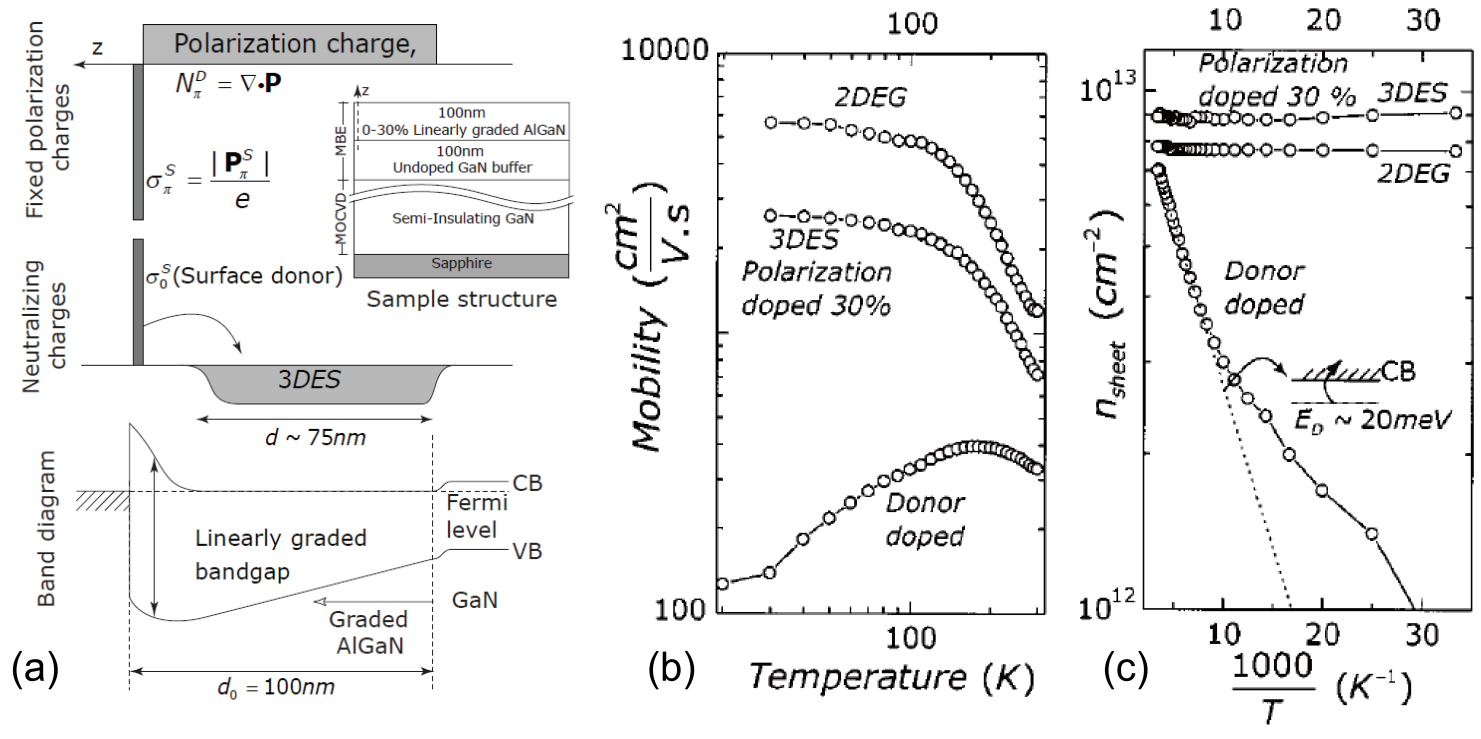

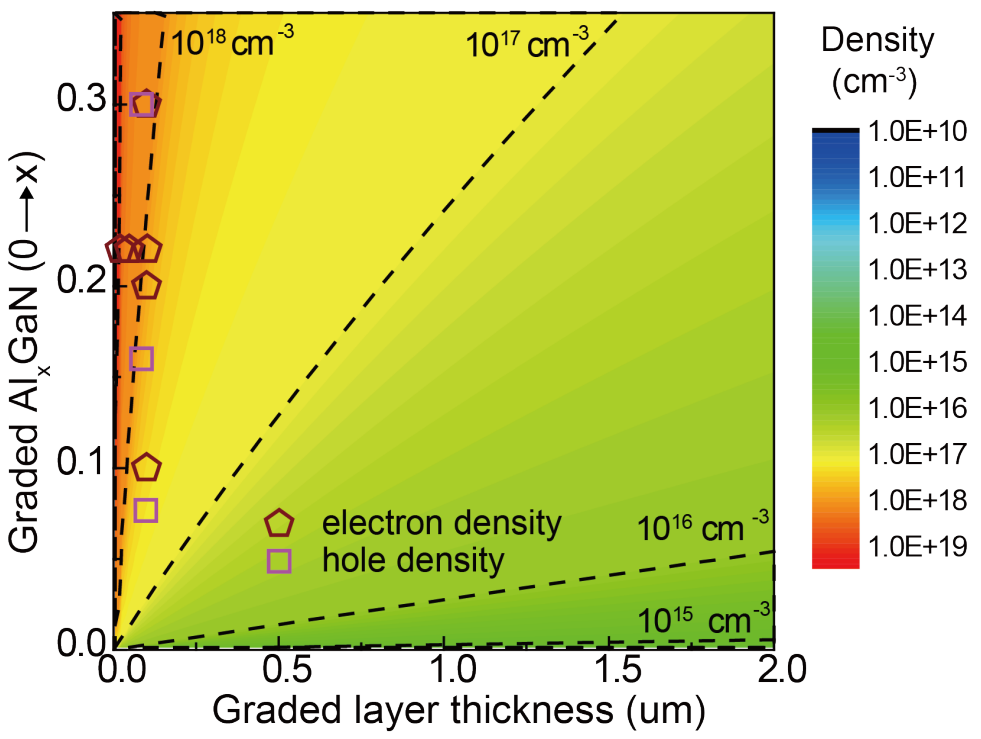

Fig.1 (a) Schematic of charge control showing polarization charges and formation of the 3DES. The band diagram shows depletion of the 3DES due to surface potential. Also shown is the epitaxial layer structure that generates the 3DES. (b) Temperature dependent carrier sheet densities and (c) mobility for a polarization doped, impurity donor doped and a 2DEG structures [1].

Fig.2 Charge density as a function of the graded Al composition ($x_{Al}$) and graded layer thickness ($d$). Reported experimental data are also included showing there is a wide open range to explore below $10^{17}$cm$^{-3}$, which is attractive for power switching applications.

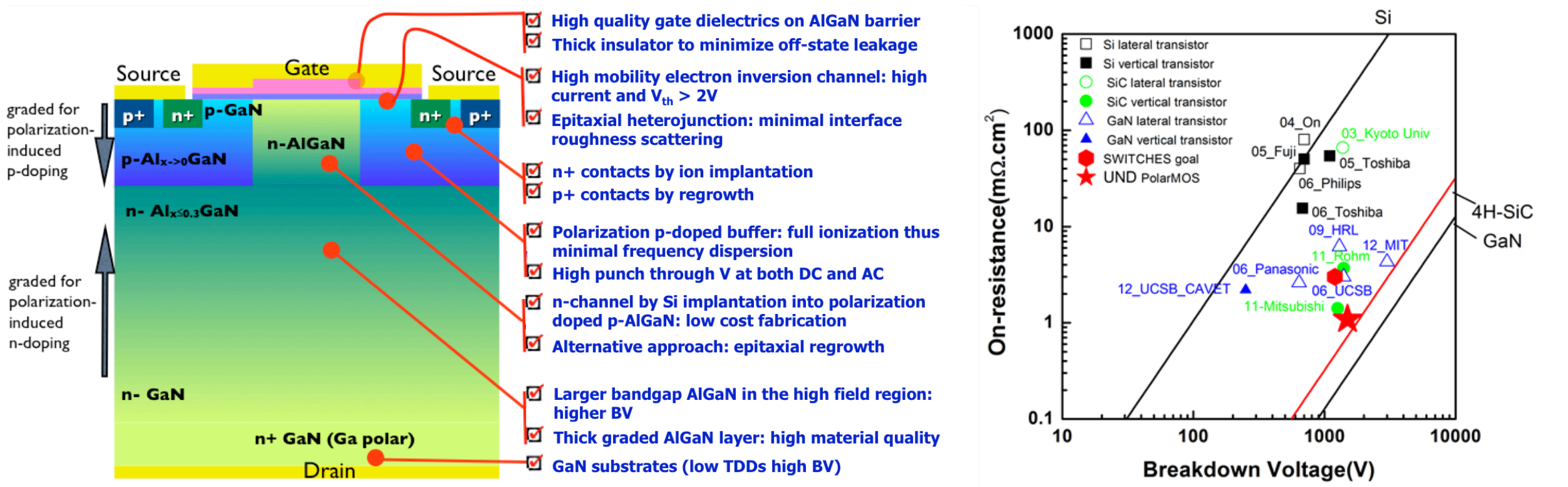

Fig.3 The ultimate GaN power transistor – PolarMOS, featuring AlGaN in the high field region to achieve higher breakdown voltage, polarization p-doped buffer to achieve full dopant ionization, thus minimizing frequency/temperature dispersion and enabling high punch through V at both DC and AC, high mobility electron inversion channel for high current and adjustable threshold voltage > 2 V. Shown on the right is the comparison of power transistors near $V_{br}$ of 1000 V. In advanced device architectures such as IGBT and superjunction MOS, Si power devices surpass the theoretical limit of unipolar devices. Both SiC and GaN power transistors exhibit room for improvement. GaN PolarMOS stands out compared to other solutions, promising a higher $V_{br}^2/R_{on,sp}$ figure of merit.

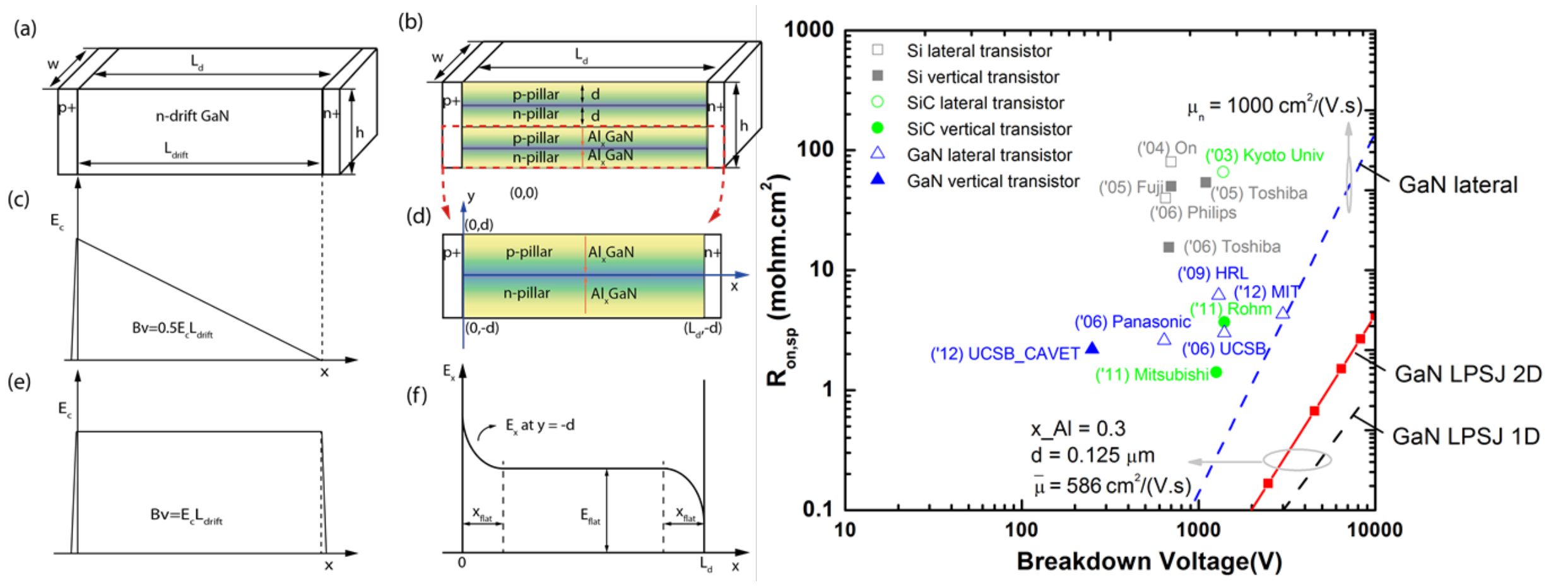

Fig.4 Schematic view of GaN (a) conventional p-n diodes and (b) lateral polarization-doped super-junction (LPSJ) diodes. One-dimensional ideal electric field distribution in (c) GaN conventional lateral p-n diodes and (e) LPSJ, and the flat field profile in LPSJ results from the balanced charge in the drift region. (d) The unit cell of LPSJ and (f) the schematic electric field distribution along the x direction at y = -d in a LPSJ showing the peak field $E_{max}$ occurs at x=0. Shown on the right is a comparison of calculated $R_{onsp}$- $V_{br}$ between the GaN conventional junction with a 1D model and LPSJ with 1D and 2D models. Also included is the reported power transistor performance near $V_{br}$ of 1000 V. The LPSJ performance based on a 2D model shows >10x reduction in $R_{on,sp}$ over conventional junctions for $BV$ > 2 kV, which is a more accurate performance estimate over the LPSJ 1D model [4,5].